# Pearltrees as a tool for referencing and teaching

**Amelia Carolina Sparavigna**

**Department of Applied Science and Technology, Politecnico di Torino, Italy**

**Abstract** Pearltrees is a social service, using which people can organise Web contents, photos and notes. These contents are the "pearls" on some trees. Besides the social opportunities, the trees can be used as a tool to organise references and for content curation. However, they can be helpful for teaching purposes too.

**Keywords:** Social Network, Reference organization, Teaching tools.

## 1. Introduction

Pearltrees is a social service which aims helping people organising Web contents, photos and notes. These are the "pearls" on some mind-map-style trees. This social service was launched in December 2010; on May 2013 it had two million monthly active users, collecting over 50,000 links every day [1]. By means of Pearltrees, users browse the internet visually by means of the "pearls", which are representing websites; by connecting these pearls, they create a network of interest. In Ref.2, it is called the "interest graph". The social networking component, allows users to follow each other and use other people pearls to build their "interest graph", collaborating to create a common tree with pearls shared among many persons. As told in Wikipedia [3], as part of the product's social features, Pearltrees users can synchronize their accounts with both Twitter and Facebook. New links added to user accounts and new Pearltrees created by users can also be broadcast via a user's Twitter and Facebook accounts, when this feature is enabled by the user.

In Ref.1, it is told that Pearltrees service might be useful for researchers, because they can organize the references (pearls) on a particular research (tree), also adding notes about them. It is also possible to highlight particular phrases or passages. Let us discuss in this paper some possible uses of this service for research and teaching purposes.

## 2. Keeping references

First of all, let us consider that the term "reference" can be used in many spheres of human knowledge, and then it can have a peculiar meaning, specific to the context in which it is used. Reference is the "act of referring" and is coming from the Medieval Latin "referential" from the verb "referre", which means "to carry back" [4]. Several words derive from this root, including referee, referent, referendum, all containing the basic meaning of Latin root as "source of origin" in terms of which something of comparable nature can be defined.

To academics, the reference is an author-title-date information in bibliographies and footnotes. A reference can be used and cited in the creation of a paper, an essay, a report or a talk. Its primary purpose is allowing people reading such work to examine the author's sources, either for validity or to learn more about the subject [5]. The references are usually listed at the end of an article or book in a specific section. Let us report a sentence from [5]: "Keeping a diary allows an individual to use references for personal organization, whether or not anyone else understands the systems of reference used. However, scholars have studied methods of reference because of their key role in communication and co-operation between different people, and also because of misunderstandings that can arise". Besides academic works, we can generalize this sentence to the searching for information that we use frequently in our life. To avoid missing reference material and reduce any inefficient use of time, Pearltrees can be an interesting tool in "keeping a diary" of relevant references on a specific subject.

Let me show an example of the use of Pearltrees for referencing: in 2011, I published a paper on ancient Green and Roman concretes (among them there are the pozzolan materials) [6]. This paper is a "pearl" in the tree created by user "jackomulso". The branch of the tree is jackomulso - dissertation - pozzolans, as shown in the Figure 1, composed by some screenshots of Pearltrees web pages. Another "pearl" is Vitruvius, because he discussed of Roman concretes in his books on the Architecture [7]. Other pearls are websites on Roman concrete. Then, using Pearltrees we can easily create a collection of pearls which are, for instance, the websites with academic references on a specific subject.

Another example of keeping references is shown in the Figure 2, composed by screenshots from the pages of user "cyborg_x1". One of the branches originating from the user's node is "hardware", and from it we arrive to the "sensors". Among the "sensors" we find the "imaging sensors", and under this node we have the references we need to know.

Let us consider that the figures we see in Pearltrees are graphs, that is, structures used to model pairwise relations between objects. A "graph" is made up of "vertices" or "nodes" and lines called "edges" that connect them. In particular, a "tree" is a graph in which any two vertices are connected by exactly one simple path (the term "tree" was coined in 1857 by the mathematician Arthur Cayley) [8]. In other words, any connected graph without simple cycles is a tree, and a "forest" is a disjoint union of trees. However, let us remark that Pearltrees is a network, not a "forest" as previously defined.

## 3. Content curation and teaching purposes

Content curation is defined as the process of collecting, organizing and displaying information relevant to a particular topic or area of interest. Then, what we have seen in the previously section, is a form of content curation, aimed to collect references for scientific papers, books and manuals. Even the reference material we find on the Web, containing useful and general information, can be considered as a form of content curation. However, let us remark that, besides a personal use, a curation service is also a procedure useful for selecting information appropriate for corporate blogs or websites, depending on the company's business sector.

The curator (again a word from a Latin verb, "curare", that is "take care") is basically the person who takes a bulk of materials and organize it. For instance, a curator or keeper of a cultural heritage institution is a content specialist responsible for an institution's collections and involved with the interpretation of heritage material [9]. Museums and galleries have then curators, who are selecting items for collection and display. Therefore, these traditional curators work with tangible objects of some sort. But recently, new kinds of curators are emerging, which are working with digital data objects. There are also curators in the world of media, for instance those people in the radio stations selecting songs to be broadcasted or in television, selecting images and videos.

Is Pearltrees a digital tool helpful for curators? The answer could be positive. Franco Torcellan, one of the users of Pearltrees, is in fact illustrating the use of Pearltrees for content curation [10]. Besides this application, he is suggesting to use of it for teaching purposes too. In particular, he is proposing Pearltrees as suitable for the study of cultural heritage and history; one of the features of this service that he is remarking is the possibility to create tree by a team of users, and then the students can actively participate with teachers to any cultural project. From the point of view of a teacher, Pearltrees can be helpful in organizing the lecture, instead of a traditional presentation by slides.

The structure of Pearltrees can be suitable for teaching sciences, such as physics, too. The trees could be created collecting "pearls" on the Web, in order to have a structure like the one we can find on HyperPhysics [11], an educational website hosted by the Georgia State University and created and maintained by Rod Nave. In fact, the information architecture of the HyperPhysics website is based on trees that organize topics from general to specific (see two screenshots in the Figure 3). In Pearltrees, I was not able to find any tree on teaching physics. However, some attempts of content curation on physics are available, as that shown in the Figure 4.

**4. Conclusion**

In this paper we have discussed a social service, Pearltrees, and shown that it can be used as a tool to organise references and for content curation. We have also proposed its use for teaching physics, instead of more traditional tools, such as the presentations by slides.

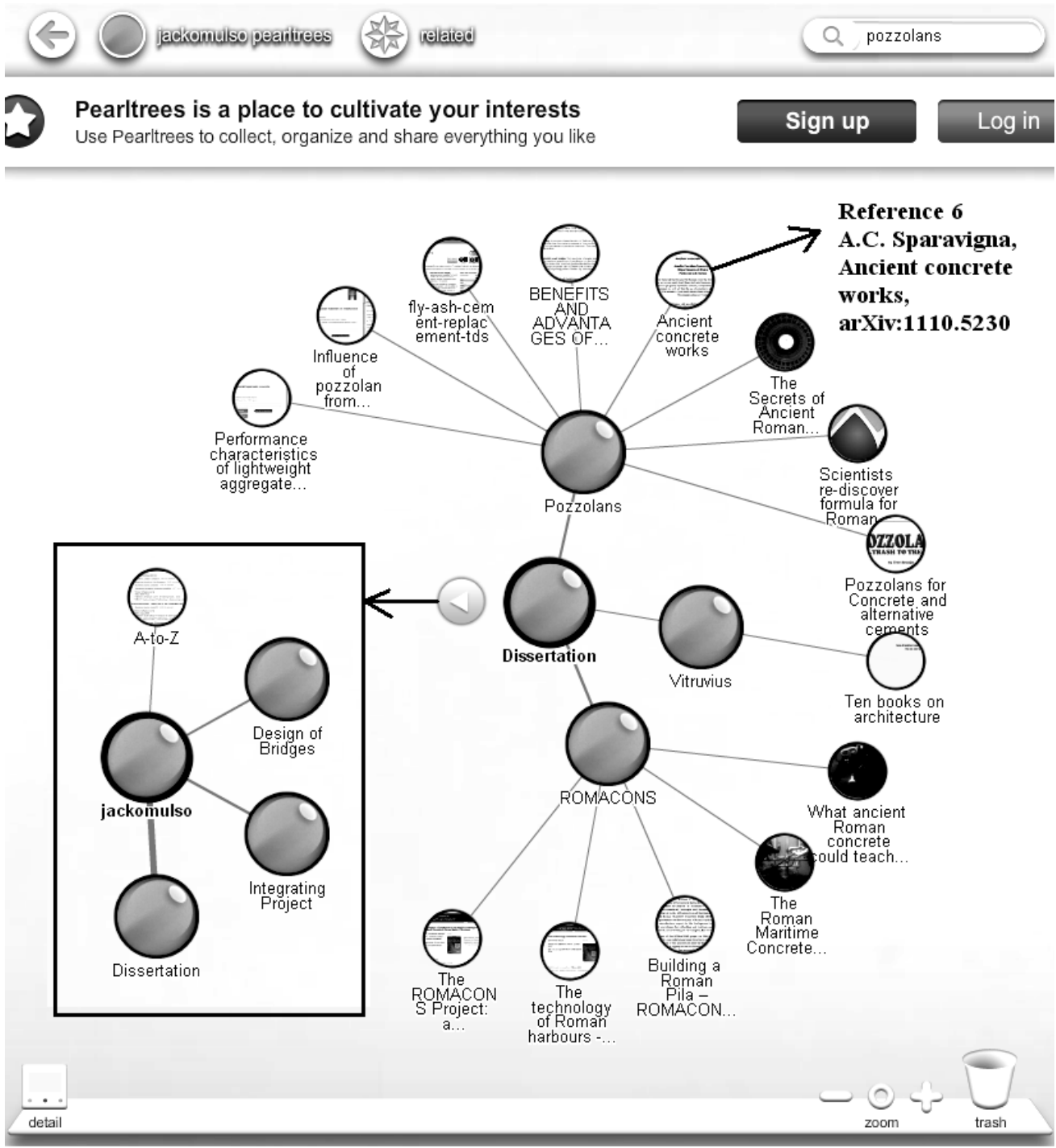

**Figure 1** – Image composed from screenshots of a user's tree, in Pearltrees.

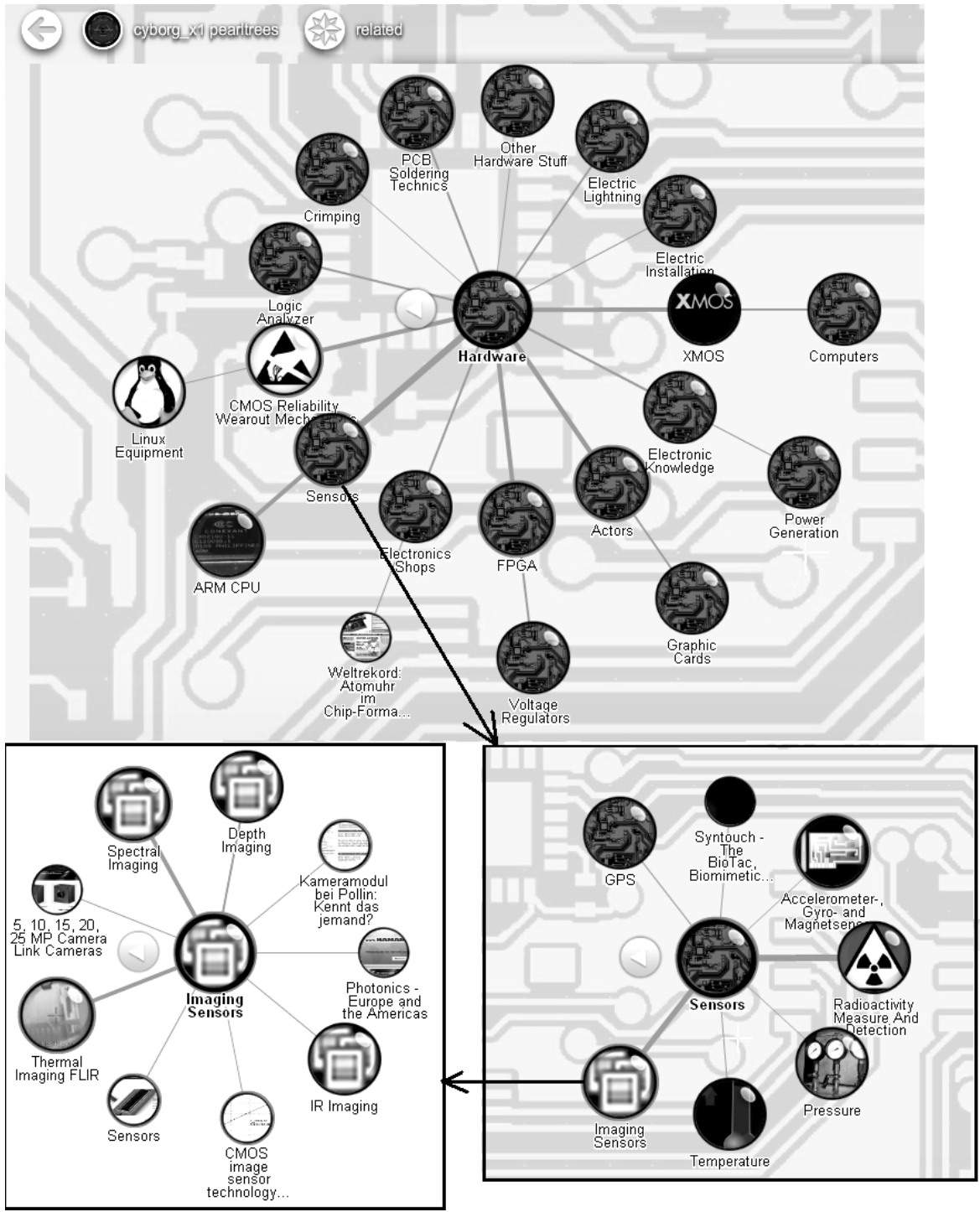

**Figure 2 – Screenshots of the cyborg_x1's tree, in Pearltrees.**

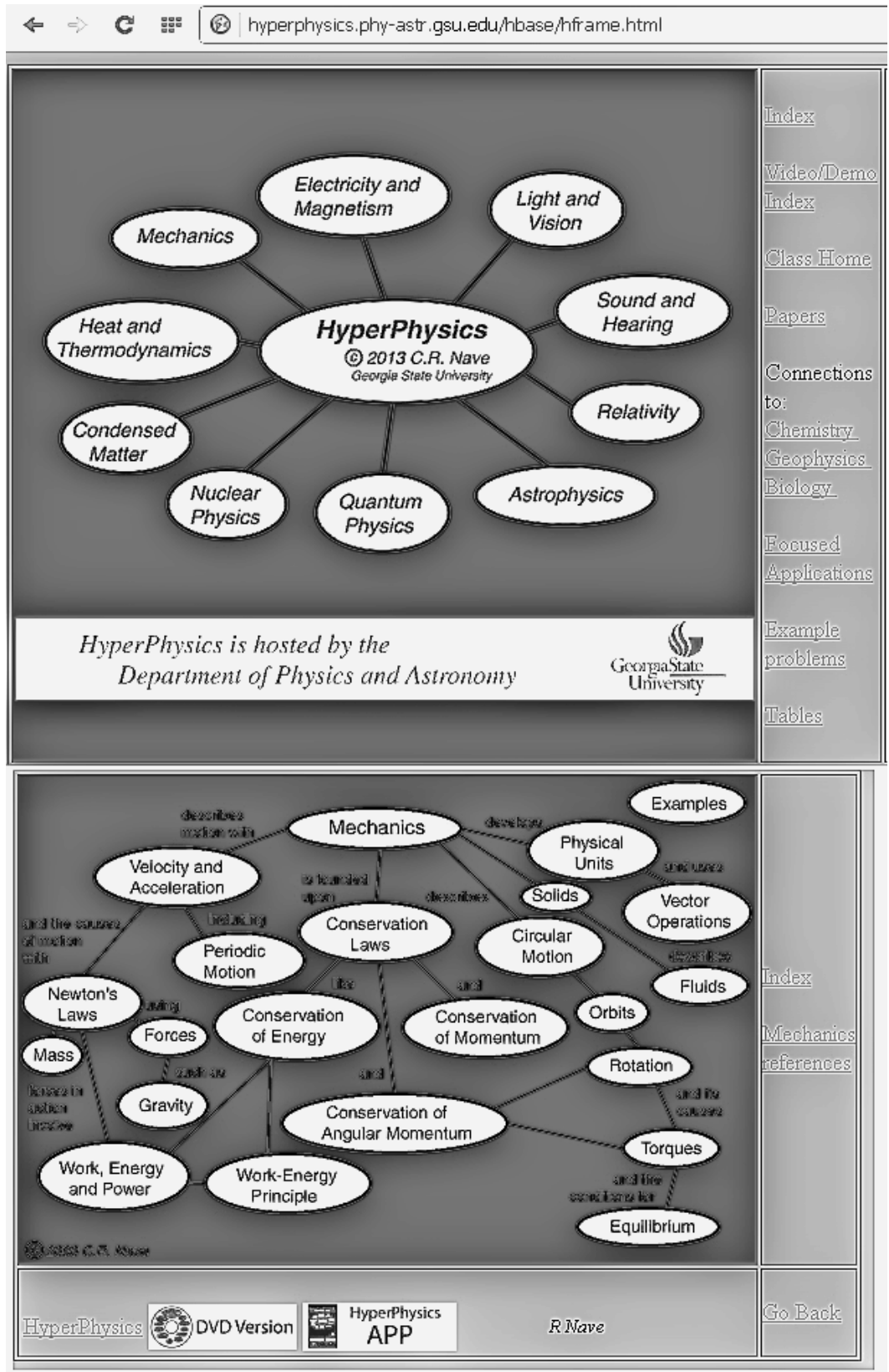

**Figure 3 - Two screenshots of Hyperphysics.**

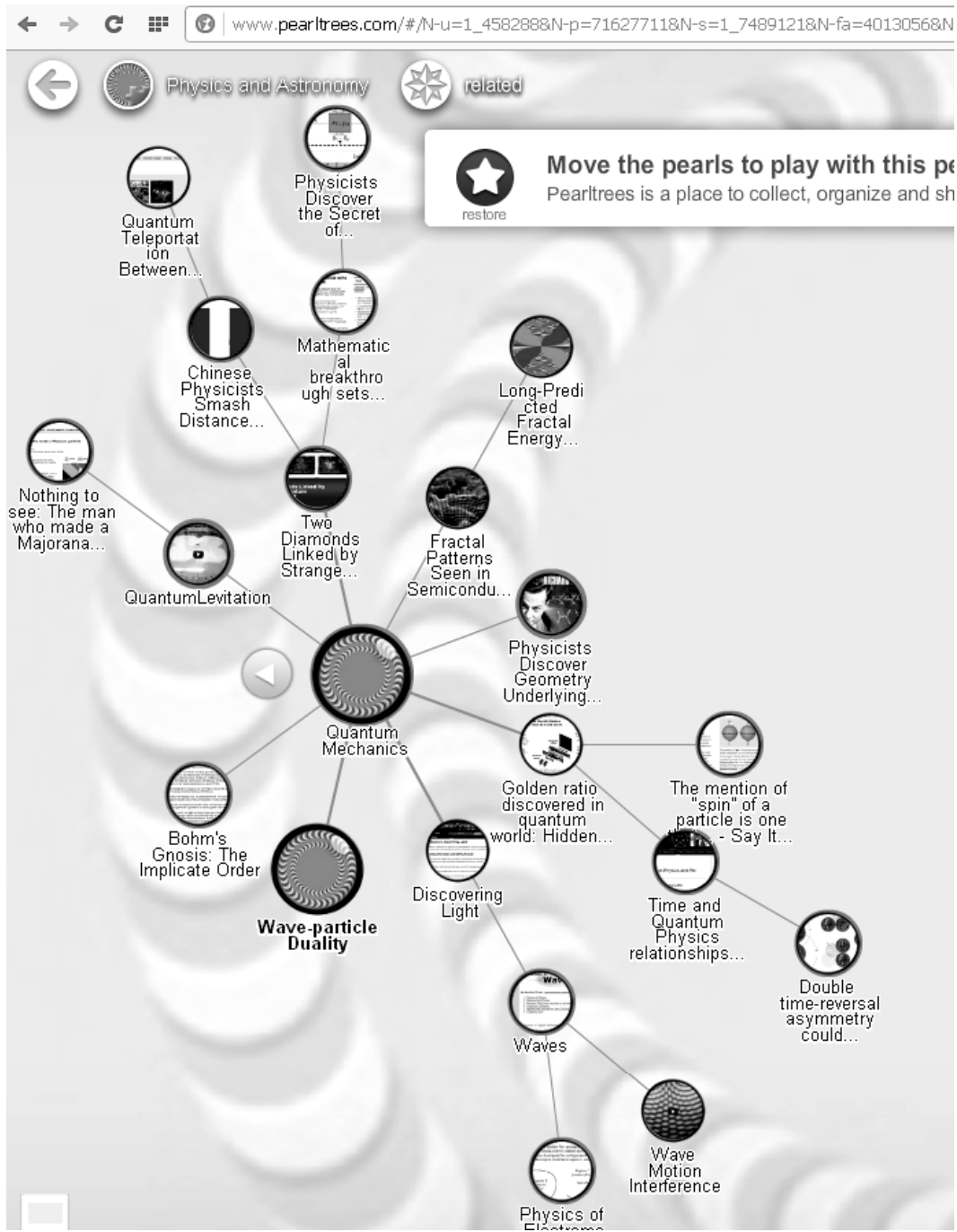

**Figure 4 – A content curation on quantum mechanics, shown by a screenshot of Physics and Astronomy, in Pearltrees.**